\documentclass{article}
\font\cero=cmss10 scaled 1728 \font\uno=cmssbx10 scaled 1200
\usepackage{tensor}
\usepackage{verbatim}
\usepackage{amsfonts}

\begin{document}
\begin{flushleft}
{\cero On the gravitational Chern-Simons action as entropy functional for three-manifolds, and the demystification of the Ho\v{r}ava-Lifshitz gravity } \\
\end{flushleft}
{\sf R. Cartas-Fuentevilla\\
{\it Instituto de F\'{\i}sica, Universidad Aut\'onoma de Puebla,
Apartado postal J-48 72570 Puebla Pue., M\'exico}.\\

We determine the more general geometrical flow in the space of metrics corresponding to the steepest descent for the three-dimensional gravitational Chern-Simons action, extending the results previously considered in  
 Class. Quantum Grav. 25 (2008) 165019,
and reveling another trouble with the four dimensional Ho\v{r}ava-Lifshitz gravity introduced in Phys. Rev. D 79, 084008 (2009), and JHEP 0903, 020 (2009), which attempts to be a candidate for an UV completion of Einstein general relativity. \\


\noindent {\uno Introduction and reconsiderations}
\vspace{.5em}

In the recent paper \cite{1} the so called Cotton flow for the metric was obtained as the gradient flow from the gravitational Chern-Simons action, which works as an entropy functional for three-manifolds. Specifically, if ${\cal F} (g)$ is that action, then its gradient along the $t$-parameter in the space of metrics is given by
\begin{equation}
     \frac{d{\cal F}}{dt} = \int_{M} \sqrt{g} C^{ij} \partial_{t} g_{ij},
     \label{CS}
\end{equation}
where $C^{ij}$ is the Cotton tensor, which is symmetric, covariantly conserved, traceless, and exclusive for three-manifolds. The following results were established in [1] from (\ref{CS}):

i) The Cotton flow, $\partial_{t} g_{ij} = C_{ij}$, leads to the steepest descent for ${\cal F}$; hence, this functional is increasing if $g$ is driven by the Cotton tensor, since $\frac{dF}{dt}= |C|^{2}$.

ii) The Cotton entropy ${\cal F} (g)$ is constant if and only if $C_{ij}$ is identically zero, and then $M$ is locally conformally flat.

iii) The stationary points $(C_{ij}=0)$ correspond to conformally flat metrics.

iv) The Yamabe flow appears in [1] only as a possible flow on the space of metrics, in general with a different parameter $\tau$, $\partial_{\tau} g_{ij} = -R g_{ij}$, and does not occur as a gradient flow of some functional entropy; in particular, such a flow seems to have no a relation with the Chern-Simons functional.

v) Under the Yamabe flow the conformal class of a metric does not change, and since the Cotton flow changes such a class, in [1] the {\cal orthogonality} of these flows is shown, by proving that $[\partial_{t}, \partial_{\tau}] g_{ij}  \sim g_{ij}$, which corresponds to another flow that does not change the conformal class.

vi) Due to the tracelessness of the Cotton tensor, the volume density of the 3-manifold is automatically preserved; hence a normalization factor is not necessary, such as in the Ricci flow scheme.

The Chern-Simons functional and its gradient are contrasted in \cite{1} with the Perelman-Hamilton approach based on the functional ${\cal F} (g,f) = \int_{M} (R+|\nabla f|^{2}) e^{-f} dV_{g}$, where the different choices of the smooth measure $e^{-f}dV_{g}$ on $M$ will correspond to different choices of diffeomorphism. If $\partial_{t}f= \partial_{t}\ln \sqrt{g}$, and $g$ is governed by the Ricci flow, then ${\cal F} (g,f)$ is an increasing functional. The comparison leads then to an additional conclusion,

vii) The Chern-Simons scheme does not require an additional scalar function $f$, as opposed to the Perelman-Hamilton scheme.

However, a simple look to the starting gradient (\ref{CS}) shows that the Cotton flow is not the more general flow leading to the steepest descent for the Chern-Simons functional:

I) Since the Cotton tensor is traceless, then the more general flow admits a corrective term proportional to the metric tensor,
\begin{equation}
     \partial_{t} g_{ij} = \eta_{0}C_{ij} + \alpha (x,t) g_{ij},
     \label{new-flow}
\end{equation}
where $\eta_{0}$ is a positive constant, and $\alpha (x,t)$ is a completely arbitrary scalar function. Under this flow the gradient of the functional, $\frac{dF}{dt}= \eta_{0} |C|^{2}$, exactly as the simple Cotton flow; hence, the functional ${\cal F}(g)$ is increasing if $g$ evolves under the generalized flow (\ref{new-flow}). This corrective term will allow to extend and change qualitatively the claims ii)-vii), and correspondingly they must be contrasted with the claims II)-VII) below.

Considering the arbitrariness of the function $\alpha$, one can choice it as a function on the curvature, for example it can be identified directly with the scalar curvature; thus, by depending on the derivatives of the metric, the $\alpha$-corrective term in the Cotton flow can change the behavior of short-wave length perturbations, and therefore affect the locally nature of that flow.

II) The Cherm-Simons entropy is not only constant if $M$ is locally conformally flat, but also for each conformal class (as expected); in Eq. (\ref{new-flow}) the choice ${\eta}_{0}=0$ (and in general $C_{ij}\neq 0$), leads to the flow $\partial_{t}g = \alpha g$, which does not change the conformal class of the metric and leaves ${\cal F}$ constant due to the tracelessness of the Cotton tensor (which in general does not vanish, and the metric class is not necessarily the conformally flat class). Such a result is certainly expected, due to the conformal invariance of the functional, this must be constant under metric deformations along the metric itself, and not only for metric locally conformally flat.

Hence, the conclusion in [1] on the constancy of the entropy functional only in its stationary points, has been generalized to metric flows within each conformal metric class; stationary points for the entropy functional do not imply necessarily stationary points for the metric flow.

III) The stationary points of the generalized flow (\ref{new-flow}) correspond to conformally flat metrics, since the constraint ${\eta}_{0}C_{ij} + \alpha g_{ij}=0$ implies that $\alpha=0$, and hence $C_{ij}=0$ (for ${\eta}_{0}\neq 0$ ), due to the tracelessness of the Cotton tensor. Thus, the flows have both the same stationary points.

IV) The Yamabe flow is intimately connected to the Chern-Simons functional; with the choice $\eta_{0}=0$ (and in general $C_{ij}\neq 0$), the generalized flow reduces to $\partial_{t}g= \alpha g$, which does not change the conformal metric class. Hence, in particular, with the choice $\alpha =-R$, the Yamabe flow is realized as a metric flow under which the functional is constant; the Yamabe flow is not only one flow of the various flows in the space of metrics, as claimed in [1].

V) The proof given in [1] on orthogonality of the Yamabe and Cotton flows, can be generalized to a general flow that does not change the conformal class, $\partial_{\tau} g_{ij} = \sigma g_{ij}$, with $\sigma$ an arbitrary function, and the generalized flow (\ref{new-flow}); the conmutator of the two flows reads
\begin{equation}
     [\partial_{t}, \partial_{\tau}] g_{ij} = (\partial_{t}\sigma - \partial_{\tau}\alpha ) g_{ij},
     \label{comm}
\end{equation}
which corresponds to another flow that preserves the conformal class. Therefore, the generalized flow (\ref{new-flow}) is a combination of orthogonal flows.

VI) In general, the volume of the 3-manifold is not preserved under the generalized flow, since $\frac{\partial V}{\partial t} = \frac{3}{2} \int_{M} \sqrt{g}\alpha d^{3}x$; due to the tracelessness of the Cotton tensor, it seems to be that one can not introduce a normalization factor, as opposed to the Ricci-flow scheme; thus, the no-preservation of the volume will be an inherent property of the steepest descent for the Chern-Simons functional.
However, this situation is apparent, since can be cured by considering more corrections for the Cotton flow (see general remarks).  

VII) In its steepest descent scenario the Chern-Simons functional contains necessarily an (completely) arbitrary scalar function, such as the Perelman-Hamilton scheme. The function $\alpha$ appears due to the traceless property of the Cotton tensor; after all, such a property is reminiscent of the conformal invariance of the Chern-Simons functional. \\

\noindent {\uno Another trouble with the Ho\v{r}ava-Lifshitz gravity}
\vspace{.5em}

The presence of the corrective $\alpha$-term has a dramatic effect on the formulation of four-dimensional 
gravity introduced in \cite{2,3}; the four-dimensional action can be split into
$S_{H}=\int dt d^{3}x\sqrt{g}N(T-V)$, where the kinetic term $T= \frac{2}{\kappa^{2}}K_{ij}G^{ijkl}K_{kl}$, 
is maintained intact under the present considerations. Furthermore,  by the detailed balance principle the superpotential $V= \frac{\kappa^{2}}{8}E^{ij}G_{ijkl}E^{kl}$, where $E$ is derived from the action principle in three-dimensions given by topologically massive gravity, which includes the Chern-Simons ${\cal F}_{CS} (g)$ plus the Einstein-Hilbert action with cosmological constant $\Lambda_{W}$ (for more details see directly \cite{2}, specifically the Eqs. (58)-(60)):
\begin{equation}
E^{ij}=\frac{1}{\sqrt{g}}\frac{\partial}{\partial g_{ij}}[ \frac{1}{w^{2}}\int{\cal F}_{CS} (g)+\mu\int\sqrt{g}(R-2\Lambda_{W})];
\end{equation}
therefore, following \cite{2} (and including of course the corrective $\alpha$-term coming from the variation of the Chern-Simons action) one obtains explicitly the action
\begin{equation}S_{H}=\int dt d^{3}x\sqrt{g}N\Big\{ (higher-derivatives)+\frac{\kappa^{2}\mu^{2}}{8(1-3\lambda)}
[\Lambda_{W}-\frac{2\alpha} {\mu w^{2} }][R-3(\Lambda_{W}-\frac{2\alpha} {\mu w^{2} })]\Big\},
\label{horava}
\end{equation}
where the short distance terms are represented by higher-derivatives terms, and do not turn out modified by the $\alpha$-term, which instead affects only the (supposed) long distance terms given by the curvature scalar and the cosmological constant; these terms combined with the kinetic term lead in \cite{2} to the result that the theory flows in the infrared  to $z=1$, limit identified then with general relativity.  We note immediately  that those conclusions are valid for a very specific gauge fixing condition, $\alpha=0$; however, as we have seen, the $\alpha$-function is completely arbitrary, and can depend on (higher) derivatives of the metric.  However, for simplicity we restrict the discussion for $\alpha$ taking only constant values; hence, in the infrared regime the theory can be compared with general relativity and leads to the emergent speed of light, Newton constant, and (effective) cosmological constant given for 
\begin{eqnarray}
c=\frac{\kappa^{2}\mu}{4}\sqrt{\frac{\Lambda_{W}-\frac{2\alpha} {\mu w^{2} }}{1-3\lambda}}, \nonumber\\
G_{N}= \frac{\kappa^{2}}{32\pi c}, \nonumber \\
\Lambda=\Lambda_{W}-\frac{2\alpha} {\mu w^{2} },\end{eqnarray}
therefore, in the scenario with $\alpha=0$, these expressions reduce to those given in \cite{2}, and the author concludes on the relevance that these effective constants emerge from a deeply nonrelativistic  theory which dominates at short distance; in this manner, the detailed balance condition will allow to connect the quantum properties of the $D+1$ dimensional theory with those of the associated theory in $D$ dimensions. However, these conclusions can be ruined with another gauge fixing condition for $\alpha$, which is suggested by the expression  adjusted of the cosmological constant:
\begin{equation}
\Lambda_{W}-\frac{2\alpha} {\mu w^{2}}=0,
\end{equation}
which leaves only the higher-derivatives terms in the action (\ref{horava}), eliminating any possibility of connecting the theory with Einstein general relativity (or its infrared modifications); in this case the emergent constants will reduce to $c=0$, $G_{N}=\infty$, and $\Lambda=0$. However, one can obtain practically what one wants, existing infinitely many possible scenarios.  In this manner, the present results will invalidate the Ho\v{r}ava-gravity based speculations on a drastic change in concepts own of general relativity such as, black holes, event horizon, information paradox, holographic principle, Bekenstein-Hawking entropy, etc., contributing to demystify a very influential theory of gravity in the last years. \\

\noindent {\uno More corrections for the Cher-Simons gradient flow}
\vspace{.5em}

As well known the Ricci flow emerges as the lowest order term in perturbative loop expansion in a two-dimensional non linear sigma model; the higher loop terms can have effects in regions where the curvature grows, for example close to singularities. Similarly, the generalized flow (\ref{new-flow}) corresponds only to the lowest order terms in  a sum of infinitely many terms that leads to the steepest descent of the Chern-Simons functional; such an infinite polynomial in the Cotton tensor can be expressed as
\begin{equation}
     \partial_{t} g_{ij} = \alpha g_{ij} + \sum^{\infty}_{s=0} \eta_{s} C^{2s+1}_{ij}, \qquad \eta_{s}\geq 0;
     \label{sum-inf}
\end{equation}
where $\eta_{s}$ are arbitrary positive constants, and the polynomial of odd order in $C$,
\begin{equation}
     C^{l}_{ij} = C_{i}{^{r}} C_{r}{^{m}} C_{m}{^{\ldots}}{_{\ldots}} C_{\ldots}{^{s}} C_{sj}; \qquad l \quad odd;
\end{equation}
under this flow the gradient of the entropy functional reduce to
\begin{equation}
     \frac{d{\cal F}}{dt} = \int_{M} \sqrt{g} d^{3}x \sum^{\infty}_{s=0} \eta_{s} | C^{s+1}_{ij}|^{2};
     \label{grad-inf}
\end{equation}
if at least one of the $\eta_{s}$ does not vanish, then the functional ${\cal F}$ is increasing.

Furthermore, considering that $C^{i}{_{j}}$ is conformal invariant and thus $\partial_{\tau} C^{i}{_{j}} =0$, one can prove that
\begin{equation}
     \partial_{\tau} \sum^{\infty}_{s=0} \eta_{s} C^{2s+1}_{ij} = \sigma \sum^{\infty}_{s=0} \eta_{s} C^{2s+1}_{ij},
     \label{sum-inf-1}
\end{equation}
and thus the commutator of the flow (\ref{sum-inf}) and the Yamabe flow reduces exactly to the same expression (\ref{comm}); therefore, the flow (\ref{sum-inf}) with infinitely many terms follows being orthogonal to the flow that does not change the conformal class.

The change of the volume along the flow (\ref{sum-inf}) is given by
\begin{equation}
     \frac{\partial V}{\partial t} =\frac{1}{2} \int_{M} \sqrt{g}d^{3}x [3\alpha + \sum^{\infty}_{s=1} \eta_{s} g^{ij} C^{2s+1}_{ij}];
     \label{grad-v}
\end{equation}
which has no a definite sign.
It should be noted nevertheless that it is not obligatory to consider the infinite sum; thus, one can consider the inclusion of specific higher terms by choosing appropriately the constants $\eta_{s}$.
Consider, for example, that $\eta_{s}=0$, except for $s=0,1$, thus we retain the lowest order terms and, $\partial_{t} g_{ij} = \alpha g_{ij} + \eta_{0}C_{ij} + \eta_{1} C_{i}{^{k}} C_{k}{^{m}}C_{m j}$; where the function $\alpha$ follows undetermined. However, the Eq. (\ref{grad-v}) allows to fix $\alpha$ with the condition of that the volume is preserved (as required for example for compact manifolds): $\alpha = -\frac{\eta_{1}}{3} g^{ij}C_{i}{^{k}} C_{k}{^{m}}C_{m j}$.

If higher order terms are included, then the stationary points spread out in the space of metrics, beyond  conformally flat manifolds (see iii), and III)); consider by simplicity the above example with $\eta_{s} \neq 0$ for $s=0,1$. The condition $\partial_{t} g_{ij}=0$ leads, by the tracelessness of $C_{ij}$, to the above expression for $\alpha$ as a cubic polynomial in $C$; thus the stationary points satisfy 
\begin{equation}
     \eta_{0} C_{ij} = \frac{\eta_{1}}{3} g^{lm} C_{l}{^{k}} C_{k}{^{r}}C_{r m} g_{ij} - \eta_{1} C_{i}{^{k}} C_{k}{^{m}}C_{m j},
     \label{st-inf}
\end{equation}
note that this constraint is satisfied in particular by conformally flat three-manifolds with $C=0$; thus in particular the Ricci-flat metrics are stationary points. However, there will be other fixed points with non-vanishing Cotton tensor, associated to different conformal classes of the metric. 

In \cite{1} the behavior of the nine homogeneous geometries under the Cotton flow was studied; four of them 
$(R^{3}, H^{3}, S^{2}\times R, H^{2}\times R)$ are fixed under that flow 
since the last three cases admit a decomposition in terms of standard metrics of the hyperbolic 3-space, the unit 2-sphere, and the hyperbolic 2-space respectively. It turns out that these geometries are stationary points for the more general flow (\ref{sum-inf}), even with the infinitely many terms included; if the sum is finite by choosing appropriately the constants $\eta_{s}$, then those geometries follow being stationary points.

In [1] were established the following conclusions: under the Cotton flow every homogeneous metric in the $SU(2)$ class converges to the round $S^{3}$ metric; every metric in the $I_{som} (R^{2})$ class evolves to a flat metric; metrics in the $Nil$ class evolve to a pancake degeneracy; metrics in the $SL(2,R)$ and $Solv$ classes develop cigar degeneracies; it would be interesting to explore the effects of the corrective terms considered in this paper on the above conclusions, especially in the cases with degeneracies.\\

\noindent {\uno General remarks}

Although we have developed explicitly the case of the Chern-Simons gradient flow given by the infinite  polynomial
(\ref{sum-inf}), a similar result can be obtained for the gradient flow for the Einstein-Hilbert action with cosmological constant, 
\begin{equation}
\partial_{t}g_{ij}= \theta_{0}(R_{ij}-\frac{1}{2}Rg_{ij}+\Lambda_{W}g_{ij})+\theta_{1}(R_{i}^{k}-\frac{1}{2}Rg_{i}^{k}+\Lambda_{W}g_{i}^{k})(R_{k}^{m}-\frac{1}{2}Rg_{k}^{m}+\Lambda_{W}g_{k}^{m})(R_{mj}-\frac{1}{2}Rg_{mj}+\Lambda_{W}g_{mj})+....,
\end{equation}
where an $"\alpha$-term" is absent, due to the Einstein-Hilbert action is not conformal invariant; the lowest order $\theta_{0}$-term is that considered traditionally,  and appears already in Eqs. (\ref{horava}). In principle these infinite polynomials are consistent with {\it detailed balance}, and can lead to uncontrollable scenarios, with a enormous number of terms and possibilities, and with a $\alpha$-function always susceptible to be re-defined.
If  {\it detailed balance} must be recovered as a principle, then must be re-considered seriously from the beginning. Finally, the present results that concern the formal structure of the Ho\v{r}ava-Lifshitz gravity, and that will presumably rule out the theory, are consistent in spirit to the phenomenological results established in \cite{4}.

\begin{center}
{\uno ACKNOWLEDGMENTS}
\end{center}
This work was supported by the Sistema Nacional de Investigadores (M\'{e}xico).


\begin{thebibliography}{}
\setlength{\itemsep}{-.50em} \setlength{\itemsep}{-.50em}
\bibitem{1} A. U. \"{O}zg\"{u}r Kisisel, \"{O}zg\"{u}r Sario\u{g}lu, and Bayram Tekin, Class. Quantum Grav. 25 (2008) 165019.
\bibitem{2} P. Ho\v{r}ava, Phys. Rev. D 79, 084008 (2009);
\bibitem{3} JHEP 0903, 020 (2009). 
\bibitem{4} C. Charmousis, G. Niz, A. Padilla and P. M. Saffin, JHEP 0908, 070 (2009).
\end{thebibliography}
\end{document}